\documentclass[a4paper,12pt]{article}
\usepackage{epsfig}
\usepackage{amssymb}
\usepackage{cite}

\newcommand{\bea}{\begin{eqnarray}}
\newcommand{\eea}{\end{eqnarray}}
\newcommand{\nm}{\nonumber\\}

\newcommand{\msLx}[1]{ \left(m^2_{\tilde{\mathcal{L}}}\right)_{#1}}
\newcommand{\snuL}[1]{ \tilde{\nu}^{\phantom{*}}_{L #1} } 
\newcommand{\snuLs}[1]{ \tilde{\nu}^{*}_{L #1} }
\newcommand{\snuLp}[1]{ \tilde{\nu}^{'}_{L #1} }
\newcommand{\snuLsp}[1]{ \tilde{\nu}^{'*}_{L #1} }
\newcommand{\snuLvev}[1]{ v_{#1} }
\newcommand{\snuLsvev}[1]{ v^*_{#1} }

\newcommand{\mux}[1]{\mu_{#1}} \newcommand{\muxs}[1]{\mu^*_{#1}}
\newcommand{\shbz}{ h_2^0 } \newcommand{\shbzs}{ h_2^{0*} }

\newcommand{\shbzvev}{ v_u }

\newcommand{\msHb}{ m^2_{H_2} }
\newcommand{\bx}[1]{b_{#1}}
\newcommand{\bxp}[1]{b_{#1}'}
\newcommand{\bxs}[1]{b^*_{#1}}

\newcommand{\MsLx}[1]{ \left({\cal{M}}^2_{\tilde{\mathcal{L}}}\right)_{#1}}
\newcommand{\MsLxb}[1]{ \left({\bf {\cal{M}}}^2_{\tilde{\mathcal{L}}}\right)_{#1}}

\newcommand{\MsLxph}[1]{ \left({\hat{\cal{M'}}}^2_{\tilde{\mathcal{L}}}\right)_{#1}}

\newcommand{\MsLxphb}[1]{ \left({{\bf\hat{\cal{M'}}}}^2_{\tilde{\mathcal{L}}}\right)_{#1}}

\newcommand{\half}{\frac{1}{2}}
\newcommand{\eps}{\epsilon}

\newcommand{\yu}{{Y}_U}
\newcommand{\lam}{{ \lambda}}
\newcommand{\lamp}{{ \lambda}^{\prime}}
\newcommand{\lampp}{{ \lambda}^{\prime\prime}}

\begin{document}
\tolerance=100000

\begin{flushright}
IPPP/05/29\\[-1mm]
DCPT/05/58\\[-1mm]
\today 
\end{flushright}

\bigskip

\begin{center}

{\Large \bf On the Neutral Scalar Sector of the General R-parity Violating
MSSM } \\[1.7cm]

{{\large A. Dedes$^{(1)}$,} 
{\large  S. Rimmer$^{(1)}$, J. Rosiek$^{(2)}$, 
M. Schmidt-Sommerfeld$^{(1)}$} }\\[0.5cm]
{\it $^{(1)}$Institute for Particle Physics Phenomenology (IPPP), Durham 
DH1 3LE, UK } \\[2mm]
{\it $^{(2)}$Institute of Theoretical Physics, Warsaw University, Hoza 69,
00-681 Warsaw, Poland } 
\\[3mm]
\end{center}

\vspace*{0.8cm}\centerline{\bf ABSTRACT}   
\vspace{0.1cm}\noindent{\small Starting out from the most general,
gauge invariant and renormalizable scalar potential of the R-parity
violating MSSM and performing a calculable rotation to the scalar
fields we arrive at a basis where the sneutrino VEVs are zero.  
The advantage of our rotation is that, in
addition, we obtain diagonal soft supersymmetry breaking sneutrino
masses and all potential parameters and VEVs real, proving that the
MSSM scalar potential does not exhibit spontaneous or explicit
CP-violation at tree level. The model has five CP-even and four CP-odd
physical neutral scalars, with at least one CP-even scalar lighter
than $M_Z$.  We parametrise the neutral scalar sector in a way that
resembles the parametrisation of the R-parity conserving MSSM, analyze
its mass spectrum, the coupling to the gauge sector and the stability
of the potential.  }

\newpage

\setcounter{equation}{0}
\section{Introduction}

That none of the terms in the Standard Model (SM) violate lepton
number (L) is not due to an imposed symmetry, but merely reflects the
fact that all such combinations of SM fields are ruled out by
consideration of gauge invariance and
renormalisability~\cite{Weinberg}.  For supersymmetric extensions of
the SM this is no longer true.  In the Minimal Supersymmetric Standard
Model (MSSM)~\cite{Reviews}, lepton number violating terms (and baryon
number (B) violating terms) appear naturally, giving rise to
tree-level processes, proton decay for example, which are already
strongly constrained by experiment.  Either, bounds can be set on
Lagrangian parameters, or a further discrete symmetry can be imposed
on the Lagrangian, such that these processes are absent.  The discrete
symmetry most commonly imposed is known as R-parity
$(R_P)$~\cite{Fayet,RpvReview}.  Under R-parity the particles of the
Standard Model including the scalar Higgs fields are even, while all
their superpartners are odd.  Imposing this symmetry has a number of
effects.  Firstly, any interaction terms which violate lepton number
or baryon number will not appear.  Secondly, the decay of the lightest
supersymmetric particle (LSP) into SM particles would violate $R_P
\,$; the LSP is therefore stable.  The sneutrino vacuum expectation
values (VEVs) are zero; without extending the MSSM field content,
spontaneous generation of $R_P$ violating terms is phenomenologically
discounted~\cite{Grossman}.

If $R_P$ conservation is not imposed, fields with different $R_P$
mix~\cite{Hall,Banks,Hempfling,Pilaftsis,Valle,Dedes,Chun}.  In
particular, the neutrinos will mix with the neutralinos and the
sneutrinos will mix with the neutral scalar Higgs fields; all five
complex neutral scalar fields can acquire vacuum expectation values.
Minimising this ten parameter potential in general is not
straightforward, it is more convenient to simplify the system by
choosing an appropriate basis in the neutral scalar sector.  As one of
the Higgs doublets carries the same quantum numbers as the lepton
doublets (apart from the non-conserved lepton number), it is
convenient to introduce the notation ${\cal{L}}_{\alpha} = (H_1,L_i)$
where $H_1$ and $L_i$ are the chiral superfields containing one Higgs
doublet and the leptons, respectively ($\alpha=0,\ldots,3$ and
$i=1,\ldots,3$).  Furthermore, starting from the interaction basis, we
are free to rotate the fields and choose the direction corresponding
to that of the ``Higgs'' field.  Assuming that the system defining the
five complex vacuum expectation values of the fields was solved, four
complex VEVs $v_{\alpha}$ would define a direction in the four
dimensional $(H_1,L_i)$ space.  One can then choose the basis vector
which defines the Higgs fields to point in the direction defined by
the vacuum expectation values.  We refer to this basis, in which the
``sneutrino'' (as we call the fields perpendicular to the ``Higgs''
field) VEVs are zero, as the vanishing sneutrino VEV basis~\cite{Banks:1995by,OK,GH}.
This basis has the virtue of simplifying the mass matrices and
vertices of the theory and thus is better suited for calculations.

Basis independent parameterisations can be chosen which explicitly
show the amount of physical lepton number
violation~\cite{Sacha2,Haber2,DH}.  Values for physical observables
such as sneutrino masses and mass splitting between CP-even and CP-odd
sneutrinos have been derived in the literature in terms of these
combinations but usually under some approximations (for example the
number of generations or CP-conservation).  We find this procedure in
general complicated for practical applications and we shall not adopt
it here.

Instead, we present in the next section a calculable procedure for
framing the most general MSSM scalar potential in the vanishing sneutrino VEV basis.  An
advantage of our procedure is to obtain a diagonal ``slepton'' mass
matrix, two real non-zero vacuum expectation values and real
parameters of the neutral scalar potential in the rotated basis.  The
latter proves that the neutral scalar sector of the most general
R-parity violating MSSM exhibits neither spontaneous nor explicit
CP-violation in agreement with~\cite{Masip}.  In Section 3, the
tree-level masses and mixing of the neutral scalar sector is
investigated.  Using the Courant-Fischer theorem for the interlaced
eigenvalues, we prove that there is always at least one neutral scalar
which is lighter than the $Z$-gauge boson. We present approximate
formulae which relate the Higgs masses, mixing and Higgs-gauge boson
vertices of the R-parity conserving (RPC) case with the R-parity
violating (RPV) one.  In Section 4, the positiveness of the scalar
mass matrices and stability of the vacuum is discussed.

\setcounter{equation}{0}
\section{Basis choice in the neutral scalar sector}

In this section we develop a procedure connecting a general neutral
scalar basis with the vanishing sneutrino VEV basis, 
the latter being more convenient for practical
applications.  The most general, renormalizable, gauge invariant
superpotential that contains the minimal content of fields, is given
by
\begin{eqnarray}
{\cal W} &=&
\eps_{ab}\left[ \frac{1}{2} \: \lam_{\alpha\beta k} \: {\cal L}_\alpha^a \, 
{\cal L}_\beta^b \,{\bar E}_k + \lamp_{\alpha j k} \: {\cal L}_\alpha^a 
\, Q_j^{b\,x} \, {\bar D}_{k\,x}  -  \mu_\alpha\: {\cal L}_\alpha^a \, H_2^b 
  +
(\yu)_{ij} \: Q_i^{a\,x} \, H_2^b \, {\bar U}_{j\,x} \right] \nonumber \\[3mm]
&+& \frac{1}{2}\, \eps_{xyz} \,\lampp_{ijk} \,{\bar U}_i^x \, {\bar
D}_j^y \, {\bar D}^z_k \;,
\label{superpot1}
\end{eqnarray}
where $Q_i^{a\,x},\;{\bar D}_i^x,\;{\bar U}_i^x,\; {\cal
L}_i^a,\;{\bar E}_i,\; H_1^a,\;H_2^a$ are the chiral superfield
particle content, $i=1,2,3$ is a generation index, $x=1,2,3$ and
$a=1,2$ are $SU(3) $ and $SU(2)$ gauge indices, respectively. The
simple form of (\ref{superpot1}) results when combining the chiral
doublet superfields with common hypercharge $Y=-\frac{1}{2}$ into
${\cal L}^a_{\alpha=0,\ldots,3}=(H_1^a,\,L_{i=1,2, 3}^a)$.
$\mu_\alpha$ is the generalized dimensionful $\mu$-parameter, and
$\lam_{\alpha \beta k},\,\lam'_{\alpha j k}, \, \lampp_{ijk}, \,
(Y_U)_{ij}$ are Yukawa matrices with $\eps_{ab}$ and $\eps_{xyz}$
being the totally anti-symmetric tensors, with
$\eps_{12}=\eps_{123}=+1$.  Then the five neutral scalar fields,
$\tilde{\nu}_{L\alpha}, h_2^0$ from the $SU(2)$ doublets, ${\cal
L}_\alpha = (\tilde{\nu}_{L\alpha}, \tilde{e}_{L\alpha}^-)^T$ and
$H_2=(h_2^+, h_2^0)^T$, form the most general neutral scalar potential
of the MSSM,
\bea
V_{\textup{\tiny{neutral}}}&=& \msLx{\alpha \beta} \snuLs{\alpha}
\snuL{\beta} + \muxs{\alpha} \mux{\beta}  \snuLs{\alpha} \snuL{\beta} 
+ \muxs{\alpha} \mux{\alpha} \shbzs \shbz + \msHb \shbzs \shbz \nm
&-& \bx{\alpha} \snuL{\alpha} \shbz - \bxs{\alpha} \snuLs{\alpha}
\shbzs + \frac{1}{8} (g^2 + g_2^2) [ \shbzs \shbz - \snuLs{\alpha}
\snuL{\alpha} ]^2 \;,
\label{eq:firstpot}
\eea
where general complex parameters $b_\alpha$, an hermitian matrix
$\msLx{\alpha \beta}$ and $m_{H_2}^2$ all arise from the supersymmetry
breaking sector of the theory.  The last term in (\ref{eq:firstpot})
originates from the D-term contributions of the superfields ${\cal
L}_\alpha, H_2$.  Defining
\bea
\MsLx{\alpha \beta} \equiv \msLx{\alpha \beta} + \muxs{\alpha} \mux{\beta}, \;\;\;\;
{\rm and} \;\;\;\;
 m_2^2 \equiv \msHb + \muxs{\alpha} \mux{\alpha} \;,
\eea 
one can rewrite the potential in~(\ref{eq:firstpot}) in a compact form
as
\bea
V_{\textup{\tiny{neutral}}} &=& \MsLx{\alpha \beta} \snuLs{\alpha}
\snuL{\beta} + m_2^2 \: \shbzs \: \shbz - (\bx{\alpha} \snuL{\alpha} \shbz + {\rm H.c})
\nonumber \\[2mm]
&+& \frac{1}{8} (g^2 + g_2^2) [ \shbzs \shbz - \snuLs{\alpha}
\snuL{\alpha} ]^2 \;. 
\label{eq:genpot}
\eea 
In order to go to the vanishing sneutrino VEV basis, we redefine the ``Higgs-sneutrino''
fields
\bea
\snuL{\alpha} \ = \ U_{\alpha\beta} \snuLp{\beta} \;,
\eea
where ${\bf U}$ is a $4\times 4$ unitary matrix 
\bea
{\bf U} \ = \ {\bf V} \: {\rm diag}(e^{i \phi_\alpha}) \: {\bf Z} \;,\label{eq2.6}
\eea 
being composed of three other matrices which we define below, ${\bf
V}$ unitary and ${\bf Z}$ real orthogonal.  The potential in the
primed basis becomes,
\bea
V_{\textup{\tiny{neutral}}} &=& \biggl [Z^{T}\MsLxphb{} Z
\biggr]_{\alpha \beta} \snuLsp{\alpha} \snuLp{\beta} + m_2^2 \: \shbzs
\: \shbz \nonumber \\[2mm] &-& \biggl [(\bxp{} Z)_{\alpha}
\snuLp{\alpha} \shbz + {\rm H.c} \biggr ] + \frac{1}{8} (g^2 + g_2^2)
\biggl ( \shbzs \shbz - \snuLsp{\alpha} \snuLp{\alpha} \biggr )^2 \;,
\label{eq:genpotp}
\eea
where
\bea
\MsLxphb{} \ =\ {\rm diag}(e^{-i \phi_\alpha})\: {\bf V}^{\dagger} \:
\MsLxb{} \: {\bf V} \: {\rm diag}(e^{i \phi_\alpha})\;,
\hspace{40pt} b^{' T} \ = \ b^{T} \: {\bf V}\: {\rm diag}(e^{i
\phi_\alpha})\; .
\label{eq27}
\eea
The unitary matrix, ${\bf V}$, is chosen such that $\MsLxphb{}$ is
real and diagonal - the hat $(\hat{\phantom{x}})$ is used to denote a
diagonal matrix. The phases $\phi_\alpha$ are chosen such that
$b_\alpha'$ is real and positive [they are equal to the phases of
$(b^T V)_{\alpha}^{*}$].  The minimisation conditions for the scalar
fields are now derived, to obtain conditions for the vacuum
expectation values,
\bea
\frac{\partial V}{ \partial \snuLsp{\alpha} }
\Bigg|_{\textup{\tiny{vac}}} &=& \biggl [ Z^{T} \MsLxph{} Z
\biggr]_{\alpha \beta} \snuLp{\beta} - (b' Z)_{\alpha} \shbzs -
\frac{1}{4} (g^2 + g_2^2) \biggl ( \shbzs \shbz - \snuLsp{\gamma}
\snuLp{\gamma} \biggr ) \snuLp{\alpha}\Big|_{\textup{\tiny{vac}}} =0
\;, \nonumber \\[3mm]
\frac{\partial V}{ \partial \shbzs }\Bigg|_{\textup{\tiny{vac}}} &=&
m_2^2 \shbz - (b' Z)_{\alpha} \snuLsp{\alpha} + \frac{1}{4} (g^2 +
g_2^2) \biggl ( \shbzs \shbz - \snuLsp{\gamma} \snuLp{\gamma} \biggr )
\shbz\Big|_{\textup{\tiny{vac}}} = 0 \;,\label{eq2.8}
\eea
where ``vac'' indicates that the fields have to be replaced by their
VEVs,
\bea
\langle \snuLp{\alpha} \rangle = \frac{\snuLvev{\alpha}}{\sqrt{2}}
\;\;,\;\;
\hspace{40pt} \langle \shbz \rangle = \frac{\shbzvev}{\sqrt{2}}
\;.\label{eq2.9}
\eea
The $U(1)_Y$ symmetry of the unbroken Lagrangian was used to set the
phase of $v_u$ to zero, however, at this stage all other vacuum
expectation values will be treated as complex variables.  By combining
Eqs.~(\ref{eq2.8},\ref{eq2.9}) we obtain

\bea
\biggl [ Z^{T} \MsLxph{} Z \biggr]_{\alpha \beta} \snuLvev{\beta} -
(b' Z)_{\alpha} \shbzvev - \frac{1}{8} (g^2 + g_2^2) ( \shbzvev^2 -
\snuLsvev{\gamma} \snuLvev{\gamma} ) \:
\snuLvev{\alpha} &=& 0 \;, \label{eq211} \\[3mm] m_2^2 \shbzvev - (b'
Z)_{\alpha} \snuLsvev{\alpha} + \frac{1}{8} (g^2 + g_2^2) ( \shbzvev^2
- \snuLsvev{\gamma} \snuLvev{\gamma} ) \: \shbzvev &=& 0 \;.
\label{eq212}
\eea
In a general basis, it is difficult to solve the above system with
respect to the VEVs without making some approximations, for example
assuming small ``sneutrino'' VEVs~\cite{Dedes}.  In order to simplify
calculations we would like to find a basis where the ``sneutrino''
VEVs vanish, $v_1=v_2=v_3=0$.  In other words, we are seeking an
orthogonal matrix $Z$, such that the following equation,
\bea
\biggl [Z^{T} \MsLxph{} Z\biggr ]_{\alpha 0} \snuLvev{0} - (b'
Z)_{\alpha} \shbzvev - \half \: M_Z^2 \: \frac{v_u^2 - v_0^2}{v_u^2 +
  v_0^2} \: \snuLvev{0} \: \delta_{0 \alpha} &=& 0 \;, 
\label{eq213}
\eea
holds.  If the above system is satisfied, then a solution with zero
``sneutrino'' VEVs exists.  The other solutions, with non-vanishing
``sneutrino'' VEVs will be discussed later.  In Eq.~(\ref{eq213}),
\bea
M_Z^2= \frac{1}{4} \left( g^2 + g_2^2 \right) \left( \shbzvev^2 +
\snuLvev{0}^2 \right) \; ,\label{eq215}
\eea
is the Z-gauge boson mass squared.  It is obvious that when $v_i=0$,
$v_0$ is real.  It is now useful to define
\bea
\tan \beta \equiv \frac{v_u}{v_0} \;.
\eea
To determine ${\bf Z}$, multiplying (\ref{eq213}) by $Z_{\gamma
\alpha}$, summing over $\alpha$ and solving for $Z_{\alpha 0}$,
yields,
\bea
Z_{\alpha 0} = \frac{ b'_{\alpha} \tan \beta }{ \MsLxph{\alpha \alpha}
 - \frac{1}{2} M_Z^2 \frac{ \tan^2\beta -1}{\tan^2\beta + 1} } \;.
 \label{eq217}
\eea
For given set of model parameters, $Z_{\alpha 0}$ depends only on
$\tan \beta$ which we can now fix by solving the orthonormality
condition,
\bea
\sum_{\alpha=0}^3 Z_{\alpha0} Z_{\alpha0} \ = \ \sum_{\alpha=0}^3
\frac{ b^{' 2}_\alpha \: \tan^2\beta}{\biggl [\MsLxph{\alpha \alpha} -
\frac{1}{2} \: M_Z^2 \; \frac{ \tan^2\beta -1}{\tan^2\beta + 1} \biggr
]^2} \ = \ 1 \;.
\label{eq:ortho}
\eea
This equation can be easily be solved numerically for any given set of
model parameters.

It is worth noting that when $b_i=0$ and using notation more typical
for this case, $b_0'\equiv m_{12}^2$, $\MsLxph{00}\equiv m_1^2$,
Eq.~(\ref{eq:ortho}) reduces to one of the standard RPC MSSM equations
for the Higgs VEVs:
\bea
m_{12}^2 v_d = v_u\left[m_1^2 - \frac{1}{8}(g^2 + g_2^2) (v_u^2 -
v_d^2) \right] \;.
\eea

For some parameter choices Eq.~(\ref{eq:ortho}) may admit multiple
solutions for $\tan\beta$.  Each of the possible $\tan\beta$ specify a
different basis, and each of these bases has one solution of the
minimisation conditions with vanishing ``sneutrino'' VEVs.  The
subtlety highlighted earlier is the following: all possible solutions
of the minimisation conditions can be found in each basis, so, in
general, each basis contains a number of extrema equal to the number
of possible solutions for $\tan\beta$.  Hence, a solution with $v_i=0$
in one basis, is a solution with $v_i \neq 0$ in another basis.  The
important point to note is that by considering all possible values of
$\tan\beta$, and selecting the value which corresponds to the deepest
minima for the solution with vanishing sneutrino VEVs, all the
solutions will have been accounted for, and the vanishing sneutrino VEV basis will have
been determined correctly.  The value of the potential at the vacuum,
in terms of $\tan\beta$ is given by
\bea 
V(\tan\beta) = -\frac{M_Z^4}{2 (g^2+g_2^2)} \: \biggl
(\frac{\tan^2\beta -1}{\tan^2\beta +1} \biggr )^2 \;.
\eea
The obvious conclusion from the equation above is that the deepest
minimum of the potential is given by the solution for $\tan\beta$ or
$\cot\beta$ which is greatest.

Knowing $\tan\beta$, one should fix $m_2^2$ using
Eqs.~(\ref{eq212},\ref{eq215}-\ref{eq217}) (again in the analogy with
RPC MSSM where $m_2^2$ is usually given in terms of $M_A, \tan\beta$).
Namely
\bea
m_2^2 \ =\ Z_{\alpha0} b'_\alpha \cot\beta - \frac{1}{2} M_Z^2 \:
\frac{\tan^2\beta-1}{\tan^2\beta +1} \;.
\label{eq:mh2}
\eea
In this way $m_2^2$ is chosen to give the correct value of the the
$Z$-boson mass.

Only the first column of the $Z$ matrix, $Z_{\alpha 0}$, is defined by
Eq.~(\ref{eq217}).  The remaining elements of ${\bf Z}$ must still be
determined.  Having fixed the first column of the matrix, the other
three columns can be chosen to be orthogonal to the first column and
to each other.  This leaves us with an $O(3)$ invariant subspace, such
that the matrix ${\bf Z}$ is given by
\bea
{\bf Z} = {\bf O} \; \left(\begin{array}{cc}
1  & 0 \\
0 &  {\bf X}_{3 \times 3}   \\
\end{array}\right) \; ,
\label{eq:zdef}
\eea
where
\bea
O = \left(\begin{array}{cccc}
Z_{00} & -\sqrt{Z_{10}^2 + Z_{20}^2 + Z_{30}^3} & 0 & 0\\
Z_{10} & {Z_{00}Z_{10}\over \sqrt{Z_{10}^2 + Z_{20}^2 + Z_{30}^3}}&
-{\sqrt{Z_{20}^2 + Z_{30}^3}\over\sqrt{Z_{10}^2 + Z_{20}^2 +
	Z_{30}^3}} & 0 \\
Z_{20} & {Z_{00}Z_{20}\over \sqrt{Z_{10}^2 + Z_{20}^2 + Z_{30}^3}}&
{Z_{10}Z_{20}\over \sqrt{Z_{20}^2 + Z_{30}^3} \sqrt{Z_{10}^2 +
Z_{20}^2 + Z_{30}^3}} & -{Z_{30}\over \sqrt{Z_{20}^2 + Z_{30}^3} }\\
Z_{30} & {Z_{00}Z_{30}\over \sqrt{Z_{10}^2 + Z_{20}^2 + Z_{30}^3}} &
{Z_{10}Z_{30}\over \sqrt{Z_{20}^2 + Z_{30}^3} \sqrt{Z_{10}^2 +
Z_{20}^2 + Z_{30}^3}} & {Z_{20}\over \sqrt{Z_{20}^2 + Z_{30}^3} } \\
\end{array}\right) \; ,
\label{eq:odef}
\eea
and ${\bf X}$ is an, as yet, undetermined $3 \times 3$ orthogonal
matrix determined by three angles.  This remaining freedom can be used
to diagonalise $\biggl [ {\bf Z}^{T} \MsLxph{} {\bf Z} \biggr]_{ij}$,
i.e. the (real symmetric) ``sneutrino'' part of the ${\bf Z}^{T}
\MsLxph{} {\bf Z}$ matrix, with entries $(\hat{{M}}^2_{\rm
\widetilde{L}})_{i}$.  We have now accomplished our aim of finding the
matrices ${\bf V}$ and ${\bf Z}$ which, after inserting into potential
of Eq.~(\ref{eq:genpotp}) and dropping the primes, reduce the scalar
potential to the form
\bea
V_{\textup{\tiny{neutral}}} &=& (M^2_{{\rm \widetilde{L}}})_{\alpha
\beta} \: \snuLs{\alpha} \: \snuL{\beta} \ + \ m_2^2 \: \shbzs \:
\shbz \ - \ \biggl [ B_{\alpha} \: \snuL{\alpha} \: \shbz \ + \ {\rm
H.c} \biggr ] \nonumber \\[3mm]
&+& \frac{1}{8} (g^2 + g_2^2) \: \biggl ( \shbzs \shbz -
\snuLs{\alpha} \snuL{\alpha} \biggr )^2 \;,
\label{eq:finalpot}
\eea
where
\bea
(M^2_{{\rm \widetilde{L}}})_{\alpha \beta} \ \equiv \ \biggl
[Z^{T}\MsLxph{} Z \biggr]_{\alpha \beta} \;\;\;\;\;{\rm and}
\;\;\;\;\; B_\alpha \ \equiv \ (\bxp{} Z)_{\alpha} \;,\label{eq.2.22}
\eea
with $\MsLxph{}$ and $\bxp{}$ given by Eq.~(\ref{eq27}).  In this
basis the matrix ${\bf M_{\rm \widetilde{L}}^2}$ adopts a particularly
simple form
\bea
(M^2_{{\rm \widetilde{L}}})_{\alpha \beta} \ = \ \left (
\begin{array}{cc} B_0 \tan\beta - \frac{1}{2} M_Z^2 \cos2\beta & B_j
\: \tan\beta \\[3mm] B_i \: \tan\beta &
(\hat{{M}}^2_{\rm \widetilde{L}})_{i} \, \delta_{ij} \end{array}
\right ) \;,
\label{eq:2.23}
\eea
where there is no sum over $i$ in the down-right part of the matrix.
Notice that we did not only succeed to self consistently go to a basis
where the sneutrino VEVs are zero, but also we managed to have the
sneutrino masses $(\hat{{M}}^2_{\rm \widetilde{L}})_{i}$ diagonal and
all the parameters of the scalar potential in Eq.~(\ref{eq:finalpot})
real.

As a byproduct of our procedure, we denote here that the potential of
Eq.~(\ref{eq:finalpot}) exhibits neither spontaneous nor explicit
CP-violation at the tree level. The latter is in agreement with the
results of Ref.~\cite{Masip} following a different method. Of course,
the parameters $\mu_\alpha$ of the superpotential and the soft
supersymmetry breaking couplings stay in general complex.
The result that the neutral scalar potential is CP invariant can also 
be seen directly from Eq.~(\ref{eq:genpot}).  
By forming the complex basis $(\snuL{\alpha},\shbzs)$ the first line of the
potential can be rewritten as a matrix; a rotation can then be performed such
that the matrix is real and diagonal.  After the rotation, the second line, 
being the contribution from D-terms, contains complex parameters
in general, but the rotation matrix can be chosen such that these 
phases are set to zero. 

A question arises when we include high order corrections to the
potential.  Then the vanishing ``sneutrino'' VEVs will be shifted to
non-zero values by tadpoles originating, for example, from the ${\cal
L} Q D$ contribution in the superpotential (\ref{superpot1}). The
``sneutrino'' VEVs maybe set back to zero by a renormalization
condition such that a counterterm for these VEVs set their one
particle irreducible (1PI) tadpole corrections to zero.

To conclude, it is worth making a remark about the sign of $B_0$.  As
is clear from the form of Eqs.~(\ref{eq.2.22},\ref{eq27},\ref{eq217}),
if $\MsLxph{\alpha \alpha} - \frac{1}{2} \: M_Z^2 \; \frac{
\tan^2\beta -1}{\tan^2\beta + 1} >0 $ for all $\alpha$, $B_0$ is
always positive in the vanishing sneutrino VEV basis.

\setcounter{equation}{0}
\section{Parametrising the neutral scalar mass matrices}
\label{sec:higgs}

The neutral scalar sector of the R-parity violating MSSM is in general
very complicated.  This is due to the fact that the scalars mix
through the lepton number violating terms proportional to $B_i$,
$M_{\rm L}^2$ and unless all of these parameters and VEVs are real one
has a $10\times 10$ matrix to consider.  However, for any given set of
model parameters, one can always perform the basis change described in
the previous section and arrive to the potential defined by
Eq.~(\ref{eq:finalpot}), with only real parameters.  Consequently, the
physical neutral scalars are, at the tree level, exact CP-eigenstates.
This implies that the neutral scalar mass matrix decouples into two
$5\times 5$ matrices, one for the CP-odd particles and one for
CP-even.  In the same manner as in the R-parity conserving MSSM, once quantum
corrections are considered, the CP invariance will generically be broken~\cite{Pilaftsis:1998pe}.

Ultimately, one would like to parametrise the scalar sector
resulting from the potential in (\ref{eq:finalpot}) with as few
parameters as possible in order to make contact with phenomenology.
These parameters in the case of the R-parity conserving MSSM are: the
physical mass of the CP-odd Higgs boson
\bea
M_A^2 \ = \  \frac{2 \: B_0}{\sin 2 \beta} \;, \label{ma2}
\eea
and $\tan\beta$.  An advantage of the form of potential in
Eq.~(\ref{eq:finalpot},\ref{eq.2.22},\ref{eq:2.23}) is that, $M_A$ and
$\tan\beta$ can still be used for parametrising the general Higgs
sector in the R-parity violating MSSM.  $M_A^2$ is the mass of the
lightest CP-odd Higgs boson in the R-parity conserving MSSM; as such,
it is used here as a parameter.  $m_A^2$ is used to denote the
physical tree-level mass of the lightest CP-odd Higgs in the R-parity
violating MSSM (the convention adopted is that masses in the RPC case,
parameters in this model, are denoted by $M$, and the masses in the
RPV model are denoted by $m$).

\subsection{CP-even neutral scalar masses and couplings}

The Lagrangian after spontaneous gauge symmetry breaking contains the
terms \bea
{\cal L} \ \supset \ - \left( \begin{array}{ccc} \mathrm{Re}\,h_0^2 &
  \mathrm{Re}\,\tilde\nu_{L0} & \mathrm{Re}\,\tilde\nu_{Li}
  \end{array} \right) \mathcal{M}^2_{\rm EVEN} \left( \begin{array}{c}
  \mathrm{Re}\,h_0^2 \\
\mathrm{Re}\,\tilde\nu_{L0} \\
\mathrm{Re}\,\tilde\nu_{Lj}
\end{array} \right) \;. 
\eea
As such, the scalar CP-even Higgs squared mass matrix becomes
\bea
 \mathcal{M}^2_{\rm EVEN} = 
\left( \begin{array}{ccc}
{ \cos^2\beta M_A^2 + \sin^2 \beta M_Z^2 } &
{ - \half \sin 2 \beta ( M_A^2 +  M_Z^2 ) } &
{ - B_j } 
\\[3mm]
{ - \half \sin 2 \beta ( M_A^2 + M_Z^2 ) } &
{ \sin^2 \beta M_A^2 + \cos^2 \beta M_Z^2 } &
{  B_j \tan\beta}
\\[3mm]
{ - B_i } &
{  B_i \tan\beta} &
{ M^2_i \delta_{ij} }
\end{array} \right) \;,  \label{even}
\eea
where
\bea 
M^2_i \  \equiv  
(\hat{{M}}^2_{\rm \widetilde{L}})_{i}
+
\half \cos2\beta M_Z^2 \;, 
\label{mi}
\eea
are in fact the sneutrino physical masses of the RPC case.  It is
important here to notice that the top-left $2\times 2$ sub-matrix is
identical to the RPC case, for which the Higgs masses are given by
\bea
M_{h,H}^2 =\frac{1}{2}\left(M_Z^2 + M_A^2\pm\sqrt{(M_Z^2 + M_A^2)^2 -4
M_A^2 M_Z^2\cos^2 2\beta }\right) \;, \label{mh}
\eea
and will be used as parameters in the RPV model.

The matrix~(\ref{even}) {\it always} has one eigenvalue which is
smaller than $M_Z^2$.  This may be proved as follows: one first
observes that the upper left $2\times 2$ submatrix of (\ref{even}),
call it A, has at least one eigenvalue smaller than or equal to
$M_Z^2$. Then using the Courant-Fischer theorem~\cite{la} of the
linear matrix algebra one proves that, for one flavour, the
eigenvalues of the $3\times 3$ matrix ${\cal M}^2_{\rm EVEN}$, are
interlaced with those of A. This means that the matrix ${\cal
M}^2_{\rm EVEN}$ with $i=1$ has at least one eigenvalue smaller or
equal than $M_Z^2$. Repeating this procedure twice, proves our
statement.
Furthermore, it is interesting to notice that in the region where
$\tan\beta \gg 1$, the eigenvector $(\sin\beta, \cos\beta, 0,0,0)^{\rm
T}$ corresponds to the eigenvalue with mass approximately $M_Z^2$.
Notice that this is the same eigenvector as in the RPC case which
corresponds to the Higgs boson which couples almost maximally to the
Z-gauge boson.

Lepton flavour violating processes have not been observed as yet and
therefore, bearing in mind cancellations, the parameters $B_i
\tan\beta$ have to be much smaller than $min(M_A^2,M_i^2)$.  To get a
rough estimate, consider the dominant contribution from neutral
scalars and neutralinos in the loop~\cite{lo},
\bea
m_\nu \sim \frac{a_{ew}}{16\pi} \frac{B^2 \tan^2\beta}{\tilde{m}^3}
\lesssim 1~{\rm eV} \;,
\eea
with $\tilde{m}=max(M_A, M_i)$ and $B\sim{\cal O}(B_i)$. This shows
that
\bea
{B_i\tan\beta\over \tilde m^2}\sim {1.2 \cdot 10^{-3}\over\sqrt{\tilde
m}}\sim 0.1\% \;.
\label{eq:bsize}
\eea
With this approximation, it is not hard to find a matrix ${\bf Z_R}$
which rotates the fields into the mass basis, such that
\bea
{\bf Z_R}^T {\bf \mathcal{M}}^2_{\rm EVEN} {\bf Z_R}= {\rm
diag}\biggl[m_{h^0}^2,m_{H^0}^2, ( m^2_{\tilde{\nu}_+ })_{i} \biggr]
\;,
\eea 
with $m_{h^0}^2$ being the lightest neutral scalar mass and
%
\bea 
Z_R = \left( \begin{array}{ccc} { \cos\alpha } & { \sin\alpha } & {
-\frac{\cos(\beta-\alpha) \cos\alpha B_j} {\cos\beta (M_j^2-M_h^2)} +
\frac{\sin(\beta-\alpha) \sin\alpha B_j}{\cos\beta (M_j^2 - M_H^2)}}
\\ {- \sin\alpha } & { \cos\alpha } & { \frac{\cos(\beta-\alpha)
\sin\alpha B_j} {\cos\beta (M_j^2-M_h^2)} + \frac{\sin(\beta-\alpha)
\cos\alpha B_j}{\cos\beta (M_j^2 - M_H^2) } } \\
\frac{\cos\beta P^h_i B_i }{\cos(\beta-\alpha)} & 
\frac{\cos\beta P^H_i B_i }{\sin(\beta-\alpha)} & 
{ \delta_{ij} }
\end{array} \right) \;,
\label{zr}
\eea
where there is no sum over $i$ and ($M_j^2, \, M_h^2, \, M_H^2$) are
defined in (\ref{mi},\ref{mh}). In addition,
\bea
\tan 2 \alpha \ = \ \tan 2 \beta \: \frac{M_A^2 + M_Z^2}{M_A^2 -
M_Z^2} \;\;\;{\rm and}\;\;\; P^{h,H}_i \ = \ \frac{M_Z^2 \cos^2{2
\beta} - M_{h,H}^2}{\cos^2\beta (M_H^2 - M_h^2) (M_i^2 - M_{h,H}^2)}
\;,
\eea
(the common convention is to choose $0\le\beta\le\pi/2$ and
$-\pi/2\le\alpha\le 0$).
The mass eigenstates of the RPV model are therefore given by
\bea
h^0 &\simeq&\cos\alpha\, \mathrm{Re}\,h_0^2 - \sin\alpha\,
\mathrm{Re}\,\tilde\nu_{L0} + \left(\frac{\cos\beta P^h_i B_i
}{\cos(\beta-\alpha)} \right) \mathrm{Re}\,\tilde\nu_{Li} \;, \\[3mm] \nonumber
H^0 &\simeq&\sin\alpha\, \mathrm{Re}\,h_0^2 + \cos\alpha\,
\mathrm{Re}\,\tilde\nu_{L0} + \left(\frac{\cos\beta P^H_i B_i
}{\sin(\beta-\alpha)} \right) \mathrm{Re}\,\tilde\nu_{Li} \;, \\[3mm] \nonumber
(\tilde{\nu}_{+})_i &\simeq& \left( - \frac{\cos(\beta-\alpha)
\cos\alpha B_j } {\cos\beta (M_j^2-M_h^2)} + \frac{\sin(\beta-\alpha)
B_j \sin\alpha B_j}{\cos\beta (M_j^2 - M_H^2)} \right)
\mathrm{Re}\,h_0^2 \\[1.5mm] \nonumber
&+& \left( \frac{\cos(\beta-\alpha) \sin\alpha B_j} {\cos\beta
(M_j^2-M_h^2)} + \frac{\sin(\beta-\alpha) \cos\alpha B_j }{\cos\beta
(M_j^2 - M_H^2)} \right) \mathrm{Re}\,\tilde\nu_{L0} +
\mathrm{Re}\,\tilde\nu_{Li} \;,
\eea
with corresponding masses,
\bea
m_{h^0}^2& \simeq & M_h^2 - {M_Z^2\cos^2 2\beta - M_h^2\over (M_H^2 -
M_h^2)\cos^2\beta}\sum_{i=1}^3 {B_i^2\over
{M}^2_{i} - M_h^2} + {\cal O}({B^4\over
M^6\cos^4\beta})\;, \\[4mm]
m_{H^0}^2& \simeq & M_H^2 + {M_Z^2\cos^2 2\beta - M_H^2\over (M_H^2 -
M_h^2)\cos^2\beta}\sum_{i=1}^3 {B_i^2\over
{M}^2_{i} - M_H^2} + {\cal O}({B^4\over
M^6\cos^4\beta}) \;, \\[4mm]
(m^2_{\tilde{\nu}_+ })_{i} &\simeq & (\hat{M}^2_{\tilde{\nu}})_{i} +
{B_i^2\over\cos^2\beta}{{M}^2_{i} - M_Z^2\cos^2
2\beta\over \biggl [{M}^4_{i} -
{M}^2_{i}\: (M_A^2 + M_Z^2) + M_A^2 \, M_Z^2 \, \cos^2 
2\beta \biggr ]}\nonumber \\[3mm] & +& {\cal O}({B^4\over M^6\cos^4\beta}) \;.
\label{snu}
\eea
The above expressions, are useful in relating the masses of the
neutral scalars in the RPC and RPV case in the valid approximation $B
\tan\beta << min(M_A^2, M_i^2)$. They are presented here for the first
time except the mass in (\ref{snu}) which agrees with Ref.\cite{GH}.
We note here that these formulae are not valid if some of the diagonal
entries in the mass matrix are closely degenerated - in such case even
small $B_i$ terms lead to the strong mixing of respective fields.
However in many types of calculations (e.g. various loop calculations)
one can still formally use such expansion - in the final result one
often gets expressions of the type $\frac{f(m_1) - f(m_2)}{m_1 - m_2}$
which have a well defined and correct limit also for degenerate
masses, even if the expansion used in the intermediate steps was, in
principle, wrong.

It is interesting to note that the rotation matrix ${\bf U}$ defined
in (\ref{eq2.6}) , although explicitly calculated in this article,
does not appear to all the neutral scalar vertices.  For example, the
vertices of the CP-even neutral scalars with the gauge bosons read
as\footnote{Note that the matrix ${\bf Z}$ defined in (\ref{eq:zdef})
has nothing to do with neither ${\bf Z_R}$ nor ${\bf Z_A}$ defined in
this section.},
\bea
{\cal L}_{\rm VVH} & = & \frac{1}{2} \: \frac{g_2 \: M_Z }{ \cos\theta_w}
  \left ( \cos \beta Z_{R2s} + \sin \beta Z_{R1s} \right ) \:
  Z^{\mu}\, Z_{\mu}\, H^0_s \nonumber \\[4mm]
&+& \frac{1}{2} \: g_2 \, M_W \left ( \cos \beta Z_{R2s} + \sin \beta
 Z_{R1s} \right ) \: W^{+ \mu} \, W^-_{\mu}\, H^0_s \;,\label{lvvh}
\eea
where $H_{s=1,...,5}^0$ are the Higgs boson fields, $h^0, H^0,
(\tilde{\nu}_+)_1, (\tilde{\nu}_+)_2, (\tilde{\nu}_+)_3 $
respectively.  From (\ref{zr}) and ${\cal L}_{\rm VVH}$ above, it is
easy to see that the light Higgs boson coupling to the vector bosons
($V=Z,W$), is proportional to $\sin(\beta-\alpha)$ as in the RPC
case\footnote{We follow the conventions of Ref.\cite{Janusz}.}.
In fact, the  coupling sum rule,  
\bea
\sum_{s=1}^5 g^2_{VVH_s^0}  = g^2_{VV\phi} \;, 
\label{vvh}
\eea
valid in the RPC case for $s=1,2$, persists also here, where
$g_{VVH_s^0}$ are the couplings appearing in (\ref{lvvh}) and
$g_{VV\phi}$ the corresponding coupling appearing in the Standard
Model.

\subsection{CP-odd neutral scalar  masses and couplings}

For the CP-odd case one finds,
\bea
{\cal L} \ \supset \ - \left( \begin{array}{ccc} \mathrm{Im}\,h_0^2 &
  \mathrm{Im}\,\tilde\nu_{L0} &  \mathrm{Im}\,\tilde\nu_{Li} \end{array} \right)
  \mathcal{M}^2_{\rm ODD} \left( \begin{array}{c} \mathrm{Im}\,h_0^2
  \\
\mathrm{Im}\,\tilde\nu_{L0} \\
\mathrm{Im}\,\tilde\nu_{Lj}
\end{array} \right) \;,
\eea
%
where the CP-odd mass matrix reads,
\bea
 \mathcal{M}^2_{\rm ODD} = 
\left( \begin{array}{ccc}
{ \cos^2 \beta M_A^2 + \xi \sin^2 \beta M_Z^2 } &
{ \half \sin 2 \beta ( M_A^2 - \xi M_Z^2 ) } &
{ B_j } 
\\[3mm]
{ \half \sin 2 \beta ( M_A^2 - \xi M_Z^2 ) } &
{ \sin^2 \beta M_A^2 + \xi \cos^2 \beta M_Z^2 } &
{  B_j \tan \beta}
\\[3mm]
{ B_i } &
{ B_i \tan \beta} &
{ M^2_{i} \delta_{ij} }
\end{array} \right) \;, \label{odd}
\eea
and $\xi$ is the gauge fixing parameter in $R_{\xi}$ gauge.  In fact,
by using an orthogonal rotation
\bea 
{\cal V} = \left( \begin{array}{ccc}
{ \sin \beta } &
{ -\cos \beta } &
{ 0 } 
\\
{ \cos } &
{ \sin \beta } &
{ 0 }
\\
{ 0 } &
{ 0 } &
{ 1 }
\end{array} \right) \;,
\eea
we can always project out the would-be Goldstone mode, of the CP-odd
scalar matrix and thus
\bea
{\cal V}^{\rm T}
\mathcal{M}^2_{\rm ODD}
{\cal V}
=
\left( \begin{array}{ccc}
{ \xi M_Z^2 } &
{ 0 } &
{ 0 } 
\\[3mm]
{ 0 } &
{ M_A^2 } &
{ \frac{B_j}{\cos \beta} }
\\[3mm]
{ 0 } &
{ \frac{B_i}{\cos \beta} } &
{ M^2_{i} \delta_{ij} }
\label{eq:rotcpodd}
\end{array} \right) \;.
\eea

Under the approximation of small bilinear RPV couplings [see
Eq.~(\ref{eq:bsize})], a solution is determined for the matrix ${\bf
Z_A}$ which rotates the fields into the mass basis, such that
\bea
{\bf Z_A}^T {\bf \mathcal{M}}^2_{\rm ODD} {\bf Z_A}= {\rm diag}\biggl
[m_{\rm G^0}^2,m_{\rm A^0}^2, (m^2_{\tilde{\nu}_{-}})_i \biggr ] \;,
\eea
\bea 
Z_A = 
\left( \begin{array}{ccc}
{ \sin\beta } & { \cos\beta } & { \frac{B_j} {M_j^2-M_A^2}} \\
{- \cos\beta } & { \sin\beta } & { \frac{B_j \tan\beta} {M_j^2-M_A^2}}
\\
0 & \frac{B_i}{\cos\beta (M_i^2-M_A^2)} & { \delta_{ij} }
\end{array} \right) \;,
\eea
with the mass eigenstates given by
\bea
G^0 &\simeq& \sin\beta\, \mathrm{Im}\,h_0^2 - \cos\beta\,
\mathrm{Im}\,\tilde\nu_{L0} \;, \nonumber\\[3mm]
A^0 &\simeq& \cos\beta\, \mathrm{Im}\,h_0^2 + \sin\beta\,
\mathrm{Im}\,\tilde\nu_{L0} + \frac{B_i}{\cos\beta (M_i^2 -M_A^2)
}\mathrm{Im}\,\tilde\nu_{Li} \;, \nonumber\\[3mm]
(\tilde{\nu}_{-})_i &\simeq& \frac{B_i} {M_j^2-M_A^2}
\mathrm{Im}\,h_0^2 + \frac{B_j \tan\beta} {M_i^2-M_A^2}
\mathrm{Im}\,\tilde\nu_{L0} + \mathrm{Im}\,\tilde\nu_{Li} \;,
\eea
with corresponding masses,
\bea
m_A^2 \ & \simeq & \ M_A^2 - {1\over\cos^2\beta}\sum_{i=1}^3 {B_i^2\over
M^2_{i} - M_A^2} \ + \ {\cal O}({B^4\over M^6\cos^4\beta}) \;, \\[3mm]
(m^2_{\tilde{\nu}_{-}})_i \ & \simeq  & \ M^2_i - {B_i^2\over\left(M_A^2 -
M^2_{i}\right)\cos^2\beta} \ + \ {\cal O}({B^4\over M^6\cos^4\beta}) \;.
\eea
The coupling of the Z gauge boson to the CP-even and CP-odd neutral
scalar fields is given by
\bea
{\cal L}_{ZHA} = \frac{-i g_2}{2 c_W} \: \left[ (p_{H^0_s} -
p_{A^0_p})_\mu \: \biggl ( \sum_{\alpha=0}^3 \, Z_{R \, (2+\alpha)s}\,
Z_{A \, (2+\alpha) p} \ - \ Z_{R \, 1s} \, Z_{A
\, 1p} \biggr ) \right] \:Z^{\mu}\,  H_s^0 \, A_p^0 \,  \; , \label{zha}
\eea
where the four momenta $p^\mu_{H^0_s}, p^\mu_{A^0_p}$ are incoming and
the fields $A^0_{p=1,...5}$ correspond to $G^0, A^0,
(\tilde{\nu}_-)_1, (\tilde{\nu}_-)_2, (\tilde{\nu}_-)_3$ respectively.
One may check that the coupling $Z-G^0-h^0$ derived from (\ref{zha})
is proportional to $\sin(\alpha -\beta)$ as it should be.

\setcounter{equation}{0}
\section{Positiveness and stability of the  scalar potential} 

\subsection{Positiveness}

In general, one should inspect whether all physical masses in the
CP-odd and CP-even sector are positive.  For that, all diagonal square
subdeterminants of mass matrices should be positive.  One can easily
check that both CP-odd and CP-even mass matrices in
(\ref{even},\ref{odd}) respectively, lead, in the rotated basis, to
the same set of conditions,
\bea
{M}^2_{i} &>& 0 \;\; {\rm with} \; i=1,2,3 \;\;\;\; {\rm and}\;\; \;\;
M_A^2 >  {1\over\cos^2\beta}\sum_{i=1}^3 {B_i^2\over
{M}^2_{i}} \; .
\label{eq:posit}
\eea
Using the form of $M_A^2$ in (\ref{ma2}), the last equation can be
rewritten in the form
\bea
B_0 >  \tan\beta \sum_{i=1}^3 {B_i^2\over
{M}^2_{i}} \;.\label{326}
\eea
Excluding some very singular mass configurations, the above conditions
are rather trivially fulfilled if one takes into account the bound of
Eq.~(\ref{eq:bsize}).

\subsection{Stability}

The question of  whether the potential is stable, i.e. bounded from below,
is far more complicated.  In most cases the quartic ($D-$)term
dominates and there is no problem.  The only exception being when the
fields follow the direction
$|h_2^0|^2=\sum_{i=0}^4|\tilde\nu_{L\,i}|^2$.  In such a case, one
should check whether the remaining part of the potential is positive
along this direction.

Denoting $R\equiv\sqrt{\sum_{i=0}^3|\tilde\nu_i|^2}$ and $h_2^0 = R
e^{-i\phi}$, where $\phi$ is a free phase, and using
Eqs.~(\ref{eq:mh2},\ref{eq:2.23},\ref{mi}), one can write down the
scalar potential along this direction in the vanishing snueutrino VEV basis as
\bea
V_{\textup{\tiny{neutral}}} &=& {B_0 \over \sin\beta\cos\beta}
\snuLs{0} \snuL{0} +
\left[M^2_{i} + B_0 \cot \beta \right] \snuLs{i} \snuL{i}\nonumber\\
&+& B_i \tan \beta \left( \snuLs{0} \snuL{i} + \snuL{0} \snuLs{i}
\right) - B_{\alpha} \left( \snuL{\alpha} h_0^2 + H.c. \right)
\nonumber\\[3mm]
&\equiv& \mathbf{\tilde\nu_L^{\dagger}\, { Q} \, \tilde\nu_L} - \left(
{\mathbf B^T \, \snuL{} \,} R e^{-i\phi} \: + \: {\rm H.c.}  \right)
\;.
\label{eq:stapot}
\eea
where the real symmetric matrix $\bf Q$ is
\bea
{\bf Q} \ = \ \left(\begin{array}{cc}
M_A^2 & B_i\tan\beta \\
B_j \tan\beta & \left[M^2_{i} + B_0 \cot \beta \right]\delta_{ij} \\
\end{array}\right)\;.
\eea
Finding the stability conditions for the potential~(\ref{eq:stapot})
is difficult, it depends on nine real variables (4 moduli and five
phases of the fields). To simplify the problem, we perform one more
field rotation to the basis in which the matrix ${\bf Q}$ is diagonal.
This can be done, in general, by numerical routines (routines where
already used in calculating the vanishing sneutrino VEV basis, and therefore, finding the
stability conditions for the general scalar potential always has to
involve some numerical analysis).  We thus define the matrix $\bf P$,
$\tilde\nu_L\rightarrow P \tilde\nu_L$, as
\bea
{\bf P^{\dagger}\, Q \, P} =\mathrm{diag}(X_0, X_1, X_2 ,X_3) \;.
\eea
In fact, ${\bf Q}$ is real, so we can choose ${\bf P}$ to be real
orthogonal.  Also, we denote $D_{\beta}\equiv B_{\alpha}
P_{\alpha\beta}$.  Obviously, the rotation ${\bf P}$ preserves 
the value of $R=|h_2^0|$.

The potential becomes:
\bea
V_{\textup{\tiny{neutral}}} &=& \sum_{\alpha=0}^3\left[ X_{\alpha}
|\snuL{\alpha}|^2 - D_{\alpha} R\left( \snuL{\alpha} e^{-i\phi} \: +\:
{\rm H.c.} \right)\right] \;,
\eea
where $X_0$ has to be positive, otherwise for $\phi=0$ along the direction
$\snuL{i}=\mathrm{Im}\, \snuL{0} = 0$ the potential
$V_{\textup{\tiny{neutral}}} = |\mathrm{Re}\,\snuL{0}|^2 [ X_{0} -
D_{0} \mathrm{sign}(\mathrm{Re}\,\snuL{0})]$ falls to $-\infty$ at
least for one direction along the $\mathrm{Re}\,\snuL{0}$ axis. 
In fact the condition on  $X_\alpha$ is $X_\alpha \ge 2 |D_\alpha|$. 
Thus our first conclusion is that the
matrix ${\bf Q}$ has to be positively defined.  One can write down
appropriate conditions in the same manner as for the scalar mass
matrices; comparing with Eq.~(\ref{eq:posit}), it can be observed that
this condition is automatically fulfilled if relation~(\ref{eq:posit})
holds.

With $X_{\alpha}$ positive, one can write down the potential as:
\bea
V_{\textup{\tiny{neutral}}} &=&\sum_{\alpha=0}^3
\left|\sqrt{X_{\alpha}} \snuL{\alpha} - {D_{\alpha}\over\sqrt{X_{\alpha}}} R
e^{i\phi}\right|^2 - R^2\sum_{\alpha=0}^3 {D_{\alpha}^2\over
X_{\alpha}} \;.
\eea
To further simplify the problem, denote $\snuL{\alpha} = u_{\alpha}
e^{i(\phi-\phi_{\alpha})}$, where $u_{\alpha}\geq 0$ are field moduli
and $\phi_{\alpha}$ are free phases.  Then
\bea
V_{\textup{\tiny{neutral}}} &=& R^2\left(\sum_{\alpha=0}^3
\left|\sqrt{X_{\alpha}} {u_{\alpha}\over R} -
{D_{\alpha}\over\sqrt{X_{\alpha}}} e^{i\phi_{\alpha}}\right|^2 -
\sum_{\alpha=0}^3 {D_{\alpha}^2\over X_{\alpha}}\right) \;,
\eea
where $R=\sqrt{\sum_{i=0}^3|\tilde\nu_i|^2} = \sqrt{\sum_{i=0}^3
u_i^2}$.  Phases $\phi_{\alpha}$ can be adjusted independently of
$u_{\alpha}$.  The worst case from the point of view of potential
stability, the smallest first term inside the parenthesis, occurs for
$D_{\alpha} e^{i\phi_{\alpha}} = |D_{\alpha}|$.  Denoting further
$\epsilon_{\alpha} = u_{\alpha}/R$, $0\leq\epsilon_{\alpha}\leq 1$,
one can reduce our initial problem to the question whether the
function
\bea
g(\epsilon_{\alpha}) &=& \sum_{\alpha=0}^3 \left|\sqrt{X_{\alpha}}
\epsilon_{\alpha} - {|D_{\alpha}|\over\sqrt{X_\alpha}} \right|^2 -
\sum_{\alpha=0}^3 {D_{\alpha}^2\over X_{\alpha}} =\sum_{\alpha=0}^3
(X_{\alpha} \epsilon_{\alpha}^2 - 2 |D_{\alpha}| \epsilon_{\alpha})
\;,
\label{eq:gdef}
\eea
depending now on four real positive parameters, is non-negative on the
unit sphere $\sum_{\alpha=0}^3 \epsilon_{\alpha}^2 = 1$. In general
such problem can be solved numerically using the method of Lagrange
multipliers. For $X_i > X_0 - D_0$, the minimum occurs for
\bea
\epsilon_\alpha = \frac{|D_\alpha|}{X_\alpha + \lambda}  \;,
\eea
where $\lambda$ can be found numerically as a root of the following
equation:
\bea
\sum_{\alpha=0}^3 \frac{D_\alpha^2}{(X_\alpha + \lambda)^2} = 1 \;.
\label{eq:lam}
\eea
For smaller $X_i$, the minimum is realized for $\epsilon_i=0$ for one or
more values of $i$ and requires analysis of various special cases.
Having found the correct minimum, to prove the stability of the
potential one needs to show that the function $g$ at the minimum is
non-negative.

As shown in Eq.~(\ref{eq:bsize}), $B_i$ terms and thus also $D_i$
terms are usually very small. In this case one can set approximate,
sufficient conditions for the stability of the potential, without
resorting to  solving  Eq.~(\ref{eq:lam}), numerically. Denote $D =
\sum_{i=1}^3 D_i^2$ and $X_{min}=min(X_1,X_2,X_3)$. Then, using the
inequality $D_i\epsilon_i\leq \sqrt{\sum_{i=1}^3 D_i^2}
\sqrt{\sum_{i=1}^3 \epsilon_i^2} = D\sqrt{1-\epsilon_0^2}$, one has
\bea
g(\epsilon_\alpha) &\geq& X_0 \epsilon_0^2 + X_{min} (1- \epsilon_0^2)
+ (X_i-X_{min})\epsilon_i^2 - 2 |D_0| \epsilon_0 - 2 D
\sqrt{1-\epsilon_0^2} \;.
\label{eq:eps}
\eea
Terms $(X_i-X_{min})\epsilon_i^2$ are always non-negative.  The worst
case being when the vector $(\epsilon_1,\epsilon_2,\epsilon_3)$ is along the
minimal $X_i$ axis, where these terms vanish.  Other terms are rotation invariant
in the 3-dimensional space $(\epsilon_1,\epsilon_2,\epsilon_3)$, so
Eq.~(\ref{eq:eps}) is equivalent to finding parameters $X_0,X_{min},
D_0, D$ for which the expression~(\ref{eq:onevar}), depending on just
one real variable, is positive:
\bea
g'(\epsilon_0) = X_0 \epsilon_0^2 + X_{min} (1- \epsilon_0^2) -2
|D_0|\epsilon_0 - 2 D \sqrt{1-\epsilon_0^2}\geq 0 \;.
\label{eq:onevar}
\eea
Analysis of~(\ref{eq:onevar}) is further simplified by one more
approximation, justified for small $D$:
\bea
g'(\epsilon_0) &\geq& X_0 \epsilon_0^2 + X_{min} (1-\epsilon_0^2) - 2
|D_0| \epsilon_0 - 2 D \;.
\label{eq:finapp}
\eea
The rhs of Eq.~(\ref{eq:finapp}) is now trivial. Following approximate
conditions for the stability of the potential can be summarized as
follows:

\vspace*{0.5cm}

\begin{tabular}{|p{2cm}lp{2cm}l|} \hline
&$X_{min}$ range &&  Stability requires \\ \hline 
&$X_{min}\geq X_0 - D_0$ &&   $X_0 \geq 2 |D_0| + 2 D$\\ 
&$0<X_{min}<X_0-D_0$ &&  $(X_0-X_{min})(X_{min}-2D)\geq D_0^2$ \\ \hline
\end{tabular}

\vspace*{0.5cm}
\noindent
Both conditions are sufficient, but not minimal - we have made some
approximations and there may be parameters which do not fall into
either of the categories above, and yet still give a stable potential.
For example, if $X_0=X_1=X_2=X_3\equiv X$, one can easily derive the
exact necessary and sufficient condition for potential stability as
$X\geq 2 \sqrt{D_0^2 + D^2}$, less strict than $X\geq 2 (|D_0| + |D|)$
which would be given by the table above.

For complementary work the reader is referred to  Ref.\cite{Steve}.

\section{Conclusions}

In this letter we present a procedure for calculating the rotation
matrix which brings the neutral scalar fields of the general R-parity
violating MSSM onto the vanishing sneutrino VEV basis where they develop $n$ zero VEVs,
with $n$ being the number of flavour generations.  In doing so, we
have made no assumption about the complexity of the parameters. We
consider the case of $n=3$ generations, but our approach immediately
applies to other cases, apart from obvious modifications of the form
of ${\bf Z}$ matrix defined in~(\ref{eq:zdef}.\ref{eq:odef}).  As a
byproduct of basis change, we prove that the tree level MSSM potential
does not exhibit any form of CP-violation, neither explicit nor
spontaneous. Consequently, the neutral scalar fields can be divided into
CP-even and CP-odd sectors with the $5\times 5$ neutral scalar squared
mass matrices, taking a very simple form with only small RPV masses
sitting on their off diagonal elements. We can thus expand along small
RPV masses and find analytic approximate formulae which relate the RPC
and the RPV neutral scalar masses.  Furthermore we also find, that in
general there is always at least one neutral scalar field with mass
lighter than $M_Z$ which couples maximally to the $Z$-gauge boson in
the case of large $\tan\beta$ and large $M_A$. Our procedure for
finding the rotation matrix ${\bf U}$ has been coded\footnote{The code
will be available from {\it
http://www.fuw.edu.pl/$\sim$rosiek/physics/rpv/scalar.html}} and is
numerically stable.

In the end, we are aiming to construct the most general MSSM quantum
field theory structure resorting neither to R-parity violation nor to
other approximations. This will be useful for examining the
phenomenology of the MSSM as a whole. The convenient choice of the
basis for the neutral sector found in this paper is a first step
towards this direction.

\vspace*{1cm}

{\bf Acknowledgments}

The authors thank Apostolos Pilaftsis for helpful discussions.
A.D. and M.S.-S. would like to thank ``The Nuffield Foundation'' for
financial support. J.R. would like to thank  IPPP for the
hospitality during his visit. His work was supported in part by the
KBN grant 2 P03B 040 24 (2003-2005). S.R. acknowledges the award of a UK
PPARC studentship.


\end{document}